# Optical Characterization of PtSi/Si by Spectroscopic Ellipsometry

**Van Long Le, Tae Jung Kim, Han Gyeol Park, Hwa Seob Kim, Chang Hyun Yoo, Hyoung Uk Kim, and Young Dong Kim**

*Nano-Optical Property Laboratory and Department of Physics, Kyung Hee University, Seoul 02447, Republic of Korea*

**Junsoo Kim, Solyee Im, Won Chul Choi, Seung Eon Moon, and Eun Soo Nam**

*IT Components and Materials Industry Technology Research Department, ETRI, Daejeon 34129, Republic of Korea*

We report optical characterization of PtSi films for thermoelectric device applications by nondestructive spectroscopic ellipsometry (SE). Pt monolayer and Pt-Si multilayer which consists of 3 pairs of Pt and Si layers were deposited on p-doped-silicon substrates by sputtering method and then rapid annealing process was done to form PtSi films through intermixing of Pt and Si atoms at the interface. Pseudodielectric function data $<\varepsilon> = <\varepsilon_1> + i<\varepsilon_2>$ of the PtSi/Si samples were obtained from 1.12 to 6.52 eV by using spectroscopic ellipsometry. Employing Tauc-Lorentz and Drude models, the dielectric function ($\varepsilon$) of PtSi films were determined. We found that the composition ratio of Pt:Si is nearly 1:1 for PtSi monolayer and we observed transitions between occupied and unoccupied states in Pt 5d states. We also observed formation of PtSi layers in Pt-Si multilayer sample. The SE results were confirmed by the transmission electron microscopy and energy dispersive X-ray spectroscopy.

# I. INTRODUCTION

For decades scientists have paid much attention to development of thermoelectric devices operated as green energy resources such as thermoelectric generators (TGs). Traditionally interesting materials for fabricating TG are $Bi_2Te_3$ and $Bi_2Se_3$ which brought encouraging results. However, supply of these materials are predicted to be limited on earth and also are known to be toxic, so developing alternative materials is strongly needed [1,2]. Platinum-silicide (PtSi) materials are well known as important material in infrared detectors [3-5], thermal imaging applications [6], and contact material in microelectronic devices such as complementary metal-oxide-semiconductors and p-channel metal-oxide-semiconductors [7,8] in the past few decades. Recently, PtSi/Si multilayer structures have drawn attention as a valuable candidate for application for thermoelectric devices [9]. The major impact of this structure is that heat transmission in PtSi/Si multilayer structure is reduced significantly due to the suppression of acoustic phonon propagation at interface between silicide and silicon. Previous theoretical studies also presents that phonon transmission is limited by an interface between two different materials [10-12]. Besides, PtSi/Si multilayer structures are based mainly on silicon that is cheaper in price and more matured in planar technology than commercial thermoelectric materials and that is the second most abundant element in the Earth's crust after oxygen [13]. Additionally, the electrical mobility of PtSi material is higher than that of Si, which is important point to improve efficiency in TGs.

In this paper, we characterized optical properties of PtSi films by spectroscopic ellipsometry (SE) which is an excellent technique for measuring the complex refractive index (or dielectric function) without need of Kramers-Kronig relations [14]. Pt monolayer and Pt/Si multilayer structure were fabricated by sputtering on p-doped Si substrate and annealed to form PtSi alloy layers. Room temperature SE measurement shows the existence of PtSi alloy layers. Transmission electron microscopy (TEM) and energy dispersive X-ray spectroscopy (EDS) measurements are also used to investigate morphology and stoichiometric composition of PtSi samples.

# II. EXPERIMENT

Sample preparation is as follows. P-doped silicon substrates are cleaned in acetone, methanol, and sulfuric peroxide mixture. After removal of native oxide overlayer on the substrate by buffered oxide etchant solution, platinum single layer 5 nm was deposited by sputtering method (sample A). Similarly, Pt-Si multilayer (sample B) is also prepared by repeated deposition of Pt and Si pair layers consecutively on the silicon substrate with 1.6 nm and 8 nm of thicknesses, respectively. Fig. 1 shows the schematic of fabrication process of samples A and B. After deposition both samples are loaded into the RTA (rapid thermal annealing) chamber for silicidation and maintained at 500°C for 5 min for formation of PtSi alloy layer by intermixing of the Pt and Si atoms through thermal migration.

The ellipsometric parameters $\Psi$ and $\Delta$ were obtained at room temperature from 1.12 to 6.52 eV using conventional SE (VASE, J. A. Woollam Co., Inc.). Here, $\tan\Psi$ and $\Delta$ are the amplitude ratio and the phase difference of the complex reflectances $r_s$ and $r_p$ for *s*- (TE-) and *p*- (TM-) polarized light, respectively. The relationship between ($\Psi$, $\Delta$) and the pseudodielectric function $<\varepsilon> = <\varepsilon_1> + i<\varepsilon_2>$ is [15]

$$\frac{r_p}{r_s} = \tan\Psi \cdot \exp(i\Delta) = \frac{\sin^2\phi - \cos\phi[<\varepsilon> - \sin^2\phi]^{1/2}}{\sin^2\phi + \cos\phi[<\varepsilon> - \sin^2\phi]^{1/2}} \tag{1}$$

where $\phi$ is the angle of incidence (AOI). The measurements were performed at $\phi = 50, 60,$ and $70°$ to obtain accurate values of the dielectric response. Morphology and composition characterizations of fabricated samples were investigated by high resolution TEM (HRTEM), scanning TEM (STEM), and EDS measurements. All were performed using a JEM-2100F (JEOL) microscope with an accelerating voltage of 200 kV.

### III. RESULTS AND DISCUSSION

To obtain the dielectric response and thickness information of the PtSi layer itself of the sample A, spectra from all the three AOIs ($\phi = 50, 60, 70°$) were fitted simultaneously using multilayer model (ambient/PtSi layer/Si-substrate). Dielectric response of PtSi was described with Tauc-Lorentz (TL) dispersion model [16] and Drude model [17]. In brief; the imaginary part of the TL dielectric function is expressed as

$$\varepsilon_{2TL}(E) = \begin{cases} \dfrac{AE_0 C(E-E_g)^2}{(E^2 - E_0^2)^2 + C^2 E^2} \dfrac{1}{E} & E > E_g, \\ 0 & E \leq E_g. \end{cases} \quad (2)$$

The subscript *TL* indicates that the model is based on the Tauc joint density of states and the Lorentz oscillator. The four fitting parameters *A*, $E_0$, *C*, and $E_g$ are the amplitude, peak position, broadening, and optical band gap, respectively, and all are in units of energy. The real part of the dielectric function ($\varepsilon_{1TL}$) is then obtained through Kramers-Kronig integration.

Fig. 2 shows the result of fitting to experimental parameters of Ψ and Δ at three AOIs, verifying that the quality of fit is excellent. Complex dielectric function from 1.12 to 6.52 eV for PtSi layer was deduced by multilayer calculation and is shown in Fig. 3. The real and imaginary parts of *ε* are dashed and solid lines, respectively. The $\varepsilon_2$ spectrum of PtSi shows three strong peaks at around 1.69, 3.03, and 5.03 eV which correspond to peaks *α*, *β*, and *γ*, respectively, in previous report [18]. Since $\varepsilon_2$ spectrum of this work is completely different from that of $Pt_2Si$ [18], we believe that annealing process allowed intermixing of the Pt and Si atoms quite well to form PtSi alloys as planned. Fig. 4 shows that how the component TL and Drude structures combine to reconstruct the $\varepsilon_2$ spectrum of PtSi. The dashed lines show the contributions of the each critical point, while the solid line is the sum of the dashed lines. In Fig. 4, it is clear to observe that contribution of Drude, feature is dominated below 1.5 eV, which is a typical metallic behavior. The *α*, *β*, and *γ* peaks above 1.5 eV region are assigned to the transitions between occupied and unoccupied states which mainly consist of Pt 5*d* states according to the reported calculation [18]. The peak energies in current work are in well agreement with 1.7, 2.95, and 4-5 eV in the calculation [18]. Also we note that the density of states of Pt 5*d* in Ref. 18 shows contributions from several overlapped states. We believe that our TL analysis with two *β* peaks of the almost same peak energies but with different broadenings might be the explanation of those overlapped states. This SE result confirms the formation of PtSi after RTA process. The thickness of sample A determined by SE analysis was 12.80 nm which was confirmed with high accuracy from HTEM image with 12.77 nm as shown in the Fig. 5. Fig. 6 is the EDS spectrum which shows that the ratio of Pt:Si of our PtSi film is 1:1, confirming the quality of the film and validity of our SE analysis.

For sample B of multilayer structure, Fig. 7 shows that appropriate number of layers was formed after annealing. The notation for each film in sample B is shown in the Fig. 1(b). It is noted that unexpected intermix layer with 1.5 nm appeared in Fig. 7. The B1 and B2 layers are grown on the sites of sputtered-Pt and Si layers, respectively, and B3 is the finally formed top layer. The EDS measurement in Fig. 8 also roughly confirms the existence of different layers in chemical components. It is interesting to note that the diffusion of Pt atoms in entire layers was observed different from original plan to have Si layers in between. Clear contrast in STEM picture seems to show the diffusions of Pt and Si were well progressed. By assuming that the bright layers in Fig. 7 are either PtSi or $Pt_2Si$, while grey layers are amorphous Si, we performed multilayer calculation to compare with the experimental data. The result is shown in Fig. 9 where experimental parameters of $\Psi$ and $\Delta$ at incident angle $55^o$ are shown in open squares and circles, respectively, and model calculation with PtSi and $Pt_2Si$ are shown by dash and dash-dot lines, respectively. The simulation result and data show significant deviation. Therefore, to get optical property of the each layer in sample B, we adopted the layer thicknesses in Fig. 7 to extract dielectric functions with TL and Drude models. Optical response of each layer is shown in Fig. 10 where only imaginary parts are shown for brevity. When compared with $\varepsilon_2$ of sample A, B3 layer has similar peaks position, suggesting that surface layer B3 has PtSi phase, which is also shown in EDS results of Fig. 8. This might be understood by simple mechanism in annealing process, Pt atoms at sites of B1 layers diffused into Si in both sides, while Pt atoms at surface B3 layer can diffuse only into the film direction (not into the air). Therefore, Pt concentration should be highest at surface B3 layer, which was also shown in EDS result in Fig. 8. For the case of B1 layers, judging from the Fig. 8, Pt concentration is lower than that of B3 with ratio of Pt:Si = 1:1.5 approximately. Energy position of peak α in B1 layer is as same as that in PtSi, while peaks β and γ of B1 layer are slightly shifted from those of PtSi. The line shape $\varepsilon_2$ of B2 layer is significantly different compared with those of B1 and B3 layers. According to Fig. 8, B2 layers have lowest Pt concentration with ratio Pt:Si = 1:2, and only one peak appeared at 1.3 eV. It shows that B2 layers do not have ordinary phase of PtSi. Based on above analysis, we note that STEM image shows clear contrast to indicate different layers but with similar gray scale to inaccurate prediction of similar chemical composition. However, SE work on optical response shows significant difference, proving that nondestructive analysis by optical method can give a reasonable analysis on embedded structures.

## CONCLUSION

We have successfully determined formation of PtSi layers based on characteristic line shapes of complex dielectric function by using nondestructive SE method. The interband transitions of electrons in Pt 5d states are observed from the absorption peaks $\alpha$, $\beta$, and $\gamma$ of PtSi. Moreover, the different Pt compositions in the layers were clearly distinguished by the changes of their complex dielectric functions. The changes of Pt concentration in PtSi structure induced the shift and disappearance of the specific peaks in dielectric function spectrum. Therefore, with SE, we could determine different chemical characteristics for each layer in multilayer structure, while STEM result revealed the existence of layers with almost same brightness only to be able to indicate thickness of layers. This nondestructive optical characterization method should be useful for physical understanding and application for the thermoelectric devices based on embedded PtSi layers.

## ACKNOWLEDGMENTS

This work was supported by ETRI R&D Program (The title of research project: "Silicide/Silicon hetero-junction structure for thermoelectric device", 15ZB1300) funded by the Government of Korea and R&D Convergence Program of NST (National Research Council of Science & Technology) of Republic of Korea.**REFERENCES**

[1] M. Schwall and B. Balke, Appl. Phys. Lett. **98**, 042106 (2011).

[2] United States Geological Survey, Mineral commodity summaries 2002. (2002). Washington, DC: U.S. Government Printing Office, p. 169. Retrieved from http://minerals.usgs.gov/minerals/pubs/mcs/2002/mcs2002.pdf

[3] T. L. Li, J. S. Park, S. D. Gunapala, E. W. Jones, H. M. Del Castillo, M. M. Weeks, and P. W. Pellegrini, IEEE Electron device letter **16**, 94 (1995).

[4] T. L. Lin, J. S. Park, T. George, E. W. Jones, R. W. Fathauer, and J. Maserjian, Appl. Phys. Lett. **62**, 3318 (1993).

[5] B. Aslan and R. Turan, Infrared Physics & Technology **43**, 85 (2002).

[6] S. Fujino, T. Miyoshi, M. Yokoh, and T. Kitahara, in *Proceedings of SPIE*, edited by I. J. Spiro (California, United States of America, August 7, 1989), Vol. 1157, p. 136.

[7] M. K. Niranjan, S. Zollner, L. Kleinman, and A. A. Demkov, Phys. Rev. B **73**, 195332 (2006).

[8] V. W. Chin, M. A. Green, and J. W. V. Storey, Solid-State Electron. **36**, 1107 (1993).

[9] W. C. Choi, D. S. Jun, S. J. Kim, M. C. Shin, and M. Y. Jang, Energy **82**, 180 (2015).

[10] S. Pettersson and G. D. Mahan, Phys. Rev. B **42**, 7386 (1990).

[11] D. G. Cahill, W. K. Ford, K. E. Goodson, G. D. Mahan, A. Majumdar, H. J. Maris, R. Merlin, and S. R. Phillpot, J. Appl. Phys. **93**, 793 (2003).

[12] S. T. Huxtable, A. R. Abramson, C. L. Tien, A. Majumdar, C. LaBounty, X. F. Fan, G. Zeng, J. E. Bowers, A. Shakouri, and E. T. Croke, Appl. Phys. Lett. **80**, 1737 (2002).

[13] C. R. Nave, Abundances of the Elements in the Earth's Crust. (2013, April 19). Retrieved from http://hyperphysics.phy-astr.gsu.edu/hbase/tables/elabund.html

[14] D. E. Aspnes and A. A. Studna, Appl. Opt. **14**, 220 (1975).7

[15] R. M. A. Azzam and N. M. Bashara, Ellipsometry and Polarized Light, North-Holland, Amsterdam (1976).

[16] G. E. Jellison and F. A. Modine, Appl. Phys. Lett. **69**, 371 (1996).

[17] H. Jujiwara and M. Kondo, Phys. Rev. B **71**, 075109 (2005).

[18] H. Bentmann, A. A. Demkov, R. Gregory, and S. Zollner, Phys. Rev. B **78**, 205302 (2008).

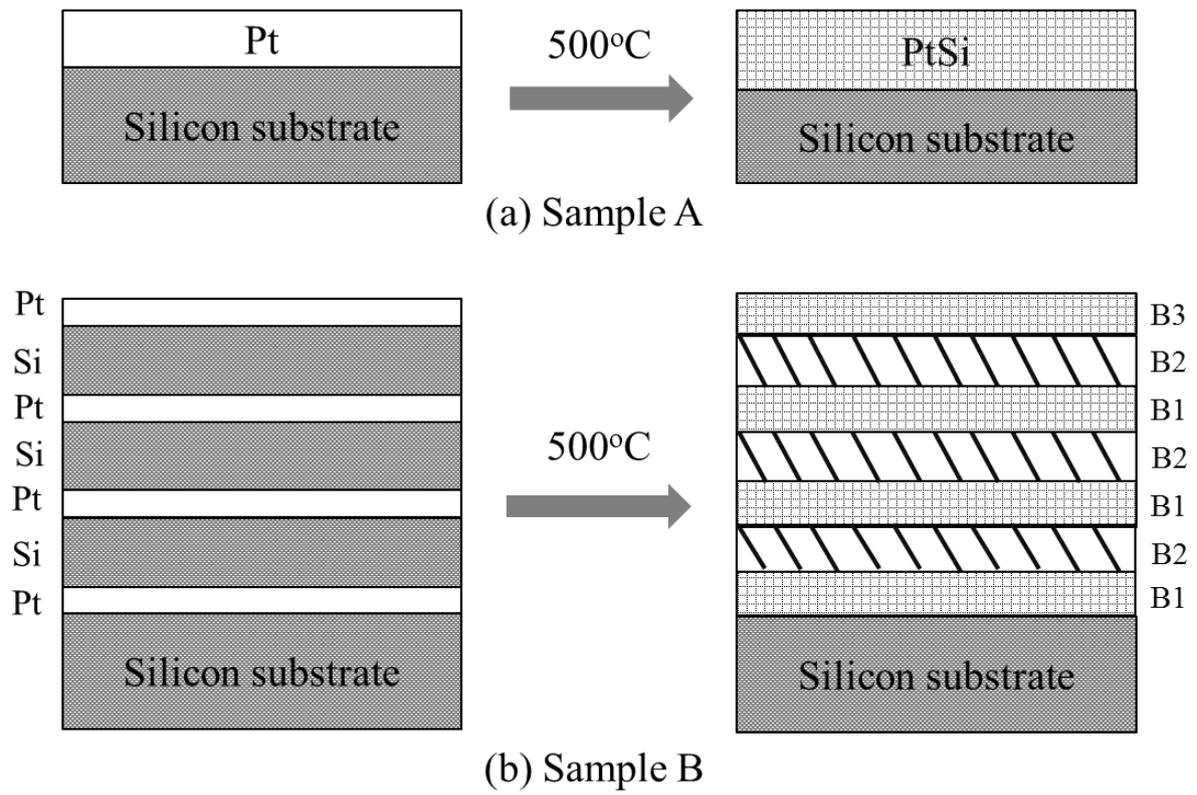

Fig. 1. Schematic structures of samples A and B before and after annealing. The blocks with the same properties are similar color.

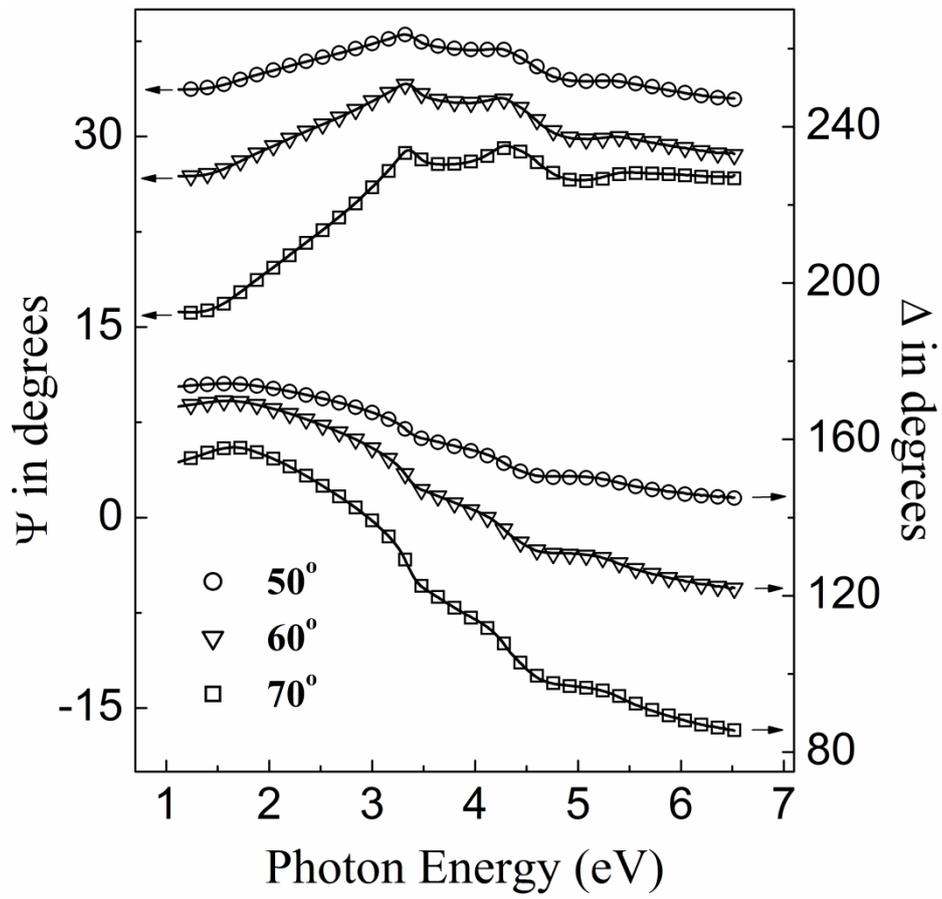

Fig. 2. Spectral magnitude (Ψ) and phase (Δ) of the ratio between *p*- and *s*-type polarization reflection coefficients, measured by ellipsometry at 50, 60, and 70° on sample A.

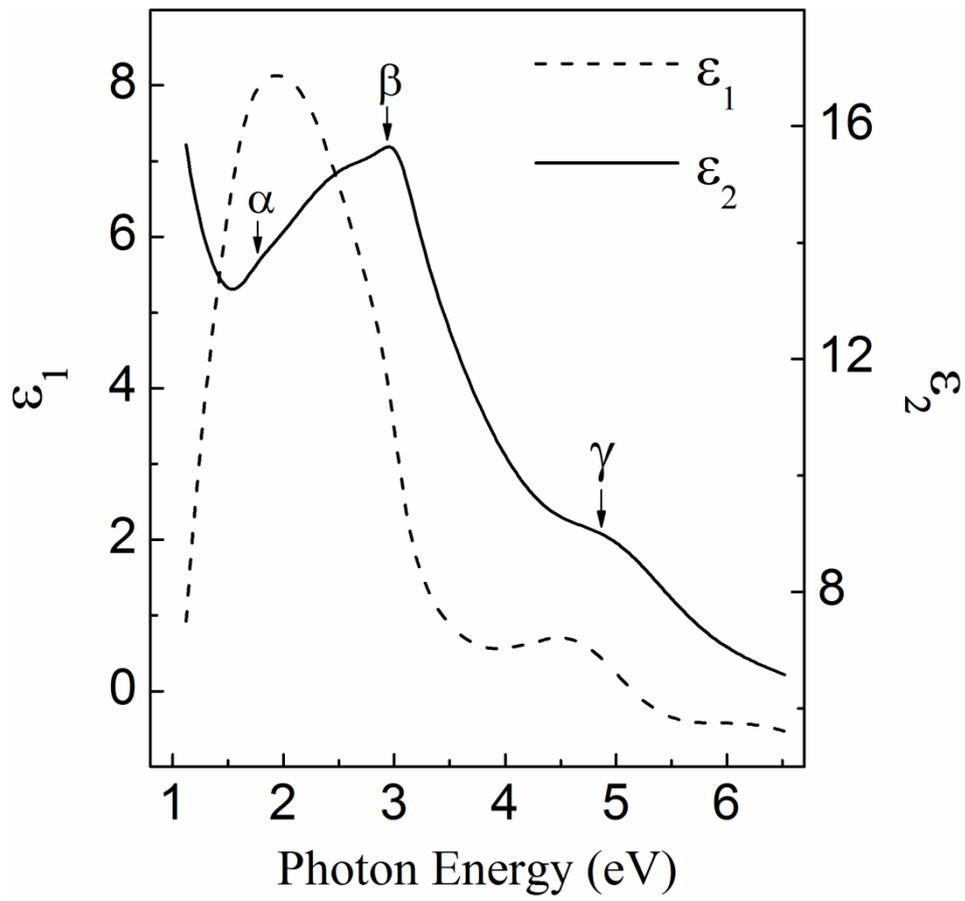

Fig. 3. Complex dielectric function determined by spectroscopic ellipsometry for sample A.

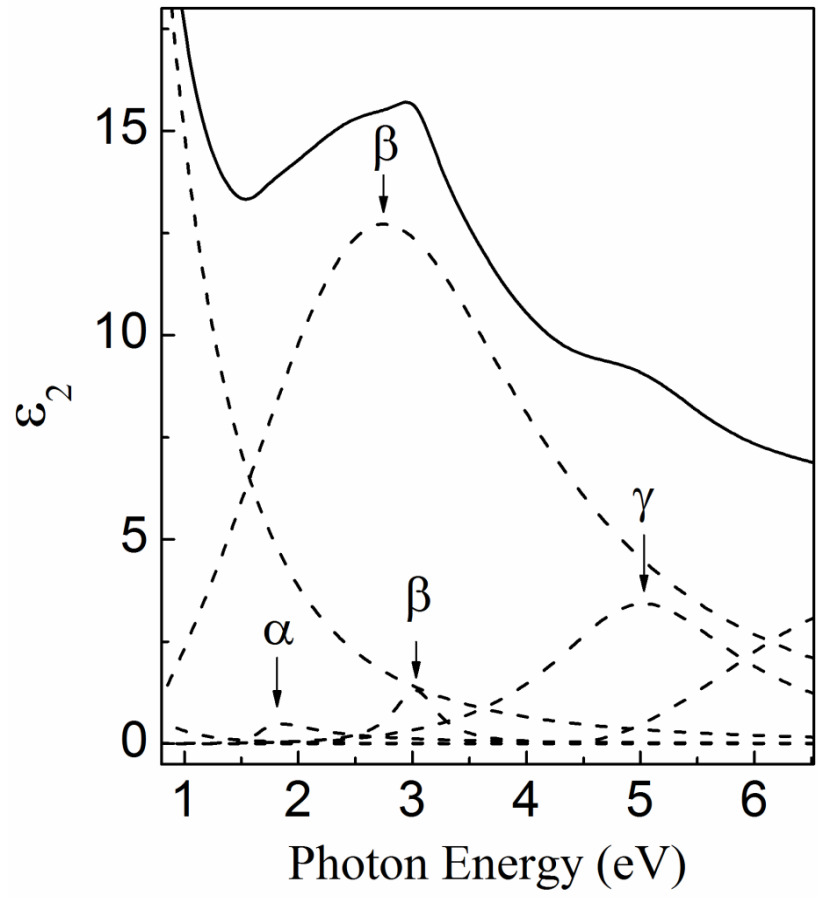

Fig. 4. $\varepsilon_2$ of PtSi is constructed from Tauc-Lorentz and Drude model.

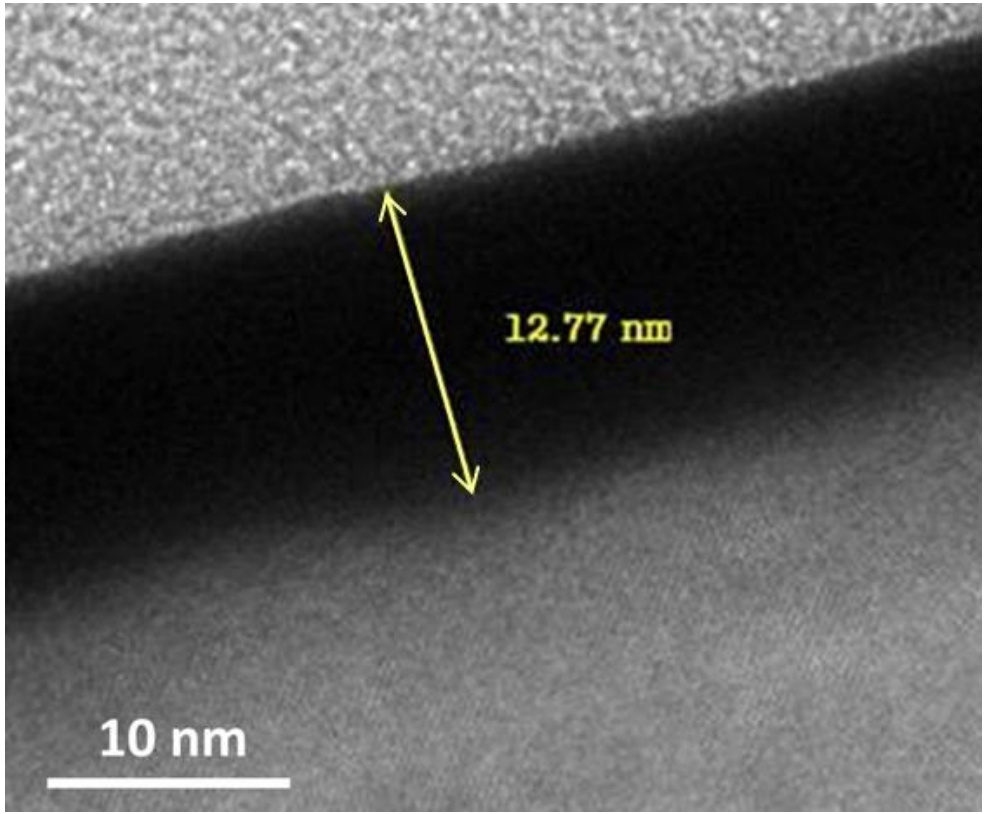

Fig. 5. High resolution TEM image of sample A.

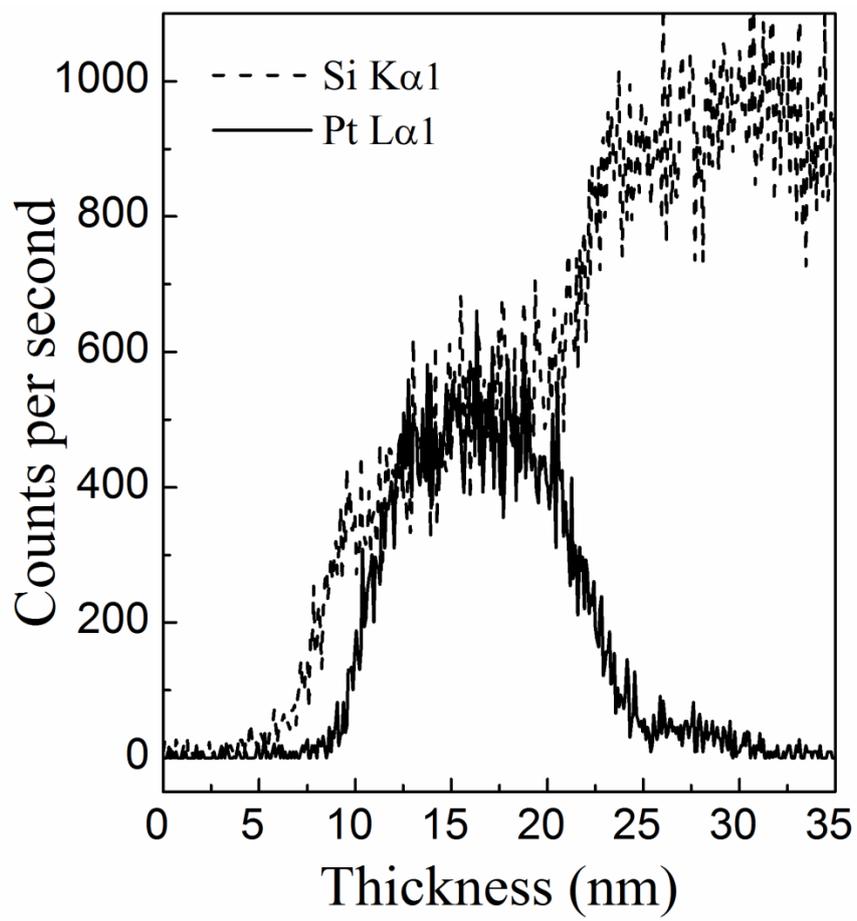

Fig. 6. EDS depth profile of sample A.

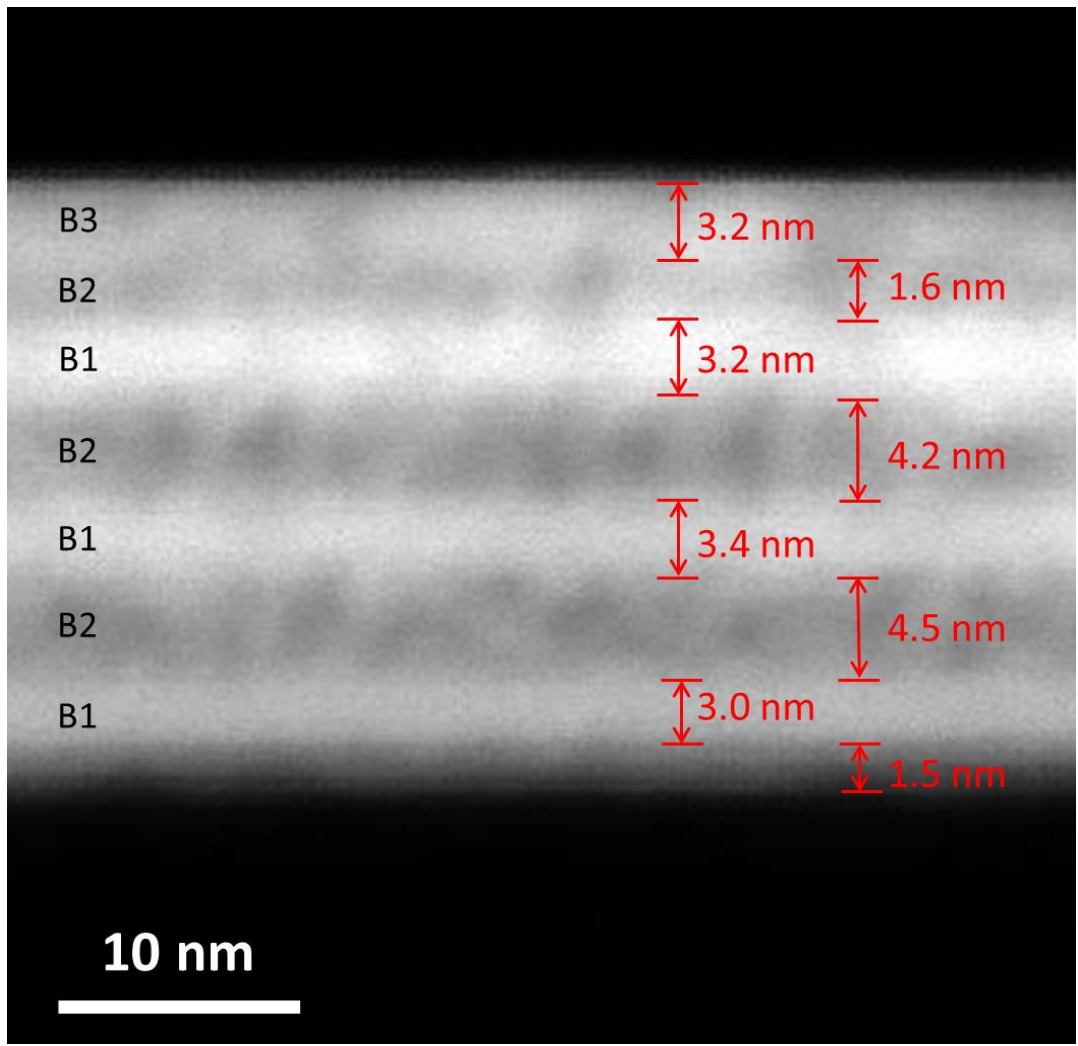

Fig. 7. STEM image of Pt/Si multilayer after annealing (sample B).

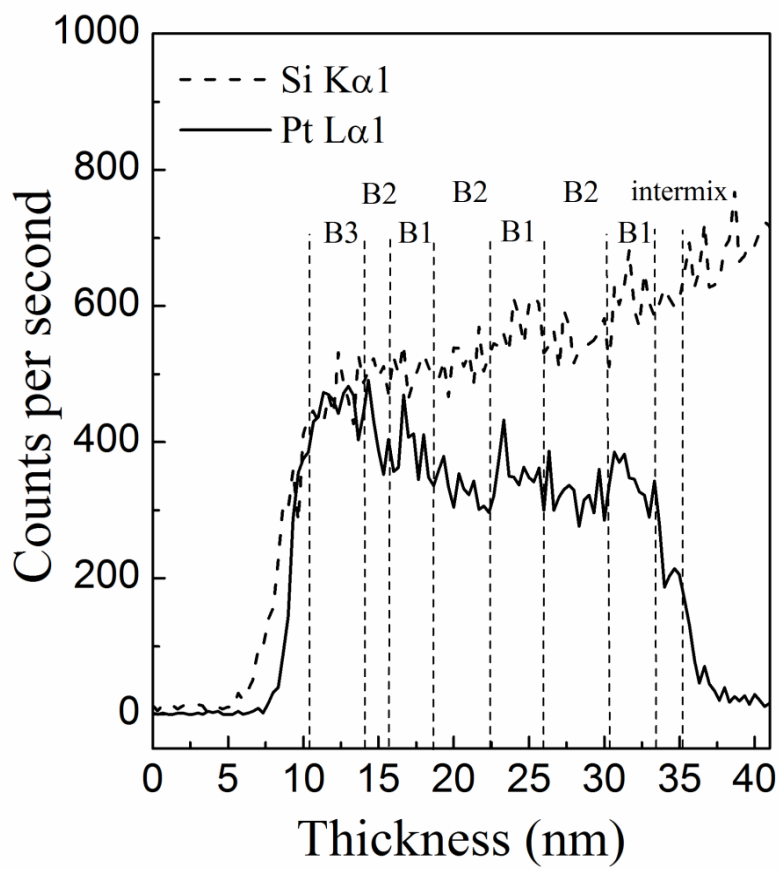

Fig. 8. EDS depth profile of sample B.

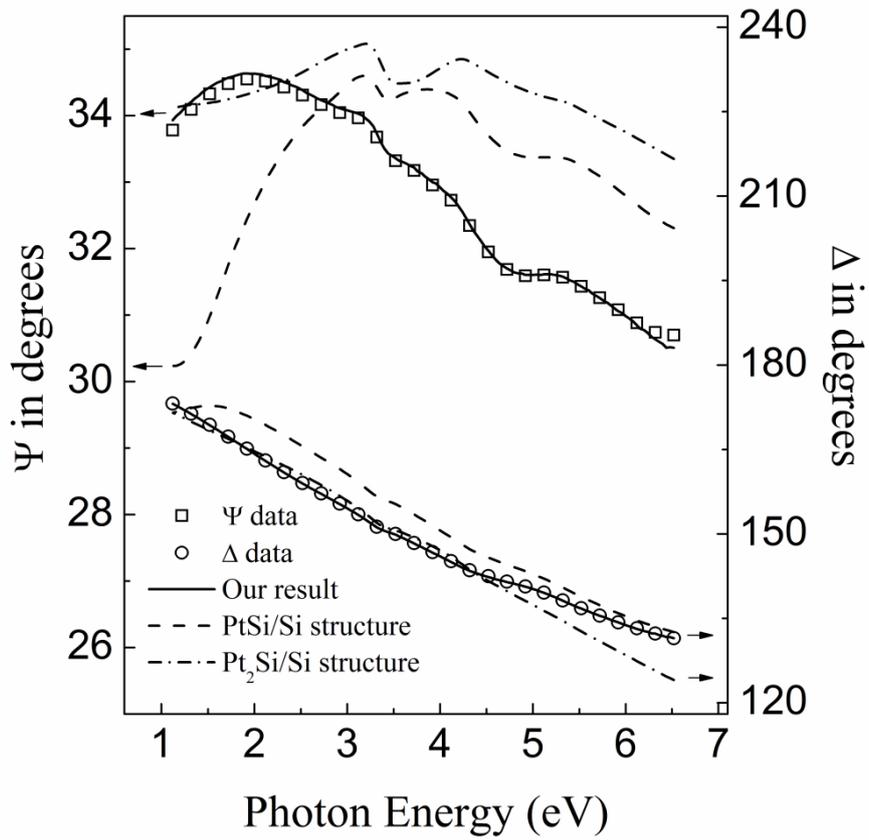

Fig. 9. Spectral magnitude (Ψ) and phase (Δ) of the ratio between *p*- and *s*-type polarization reflection coefficients, measured by ellipsometry at 55° on sample B. Solid line shows our best fitting while dashed and dash-dot lines show calculation based on PtSi/ amorphous Si and $Pt_2Si$/ amorphous Si, respectively.

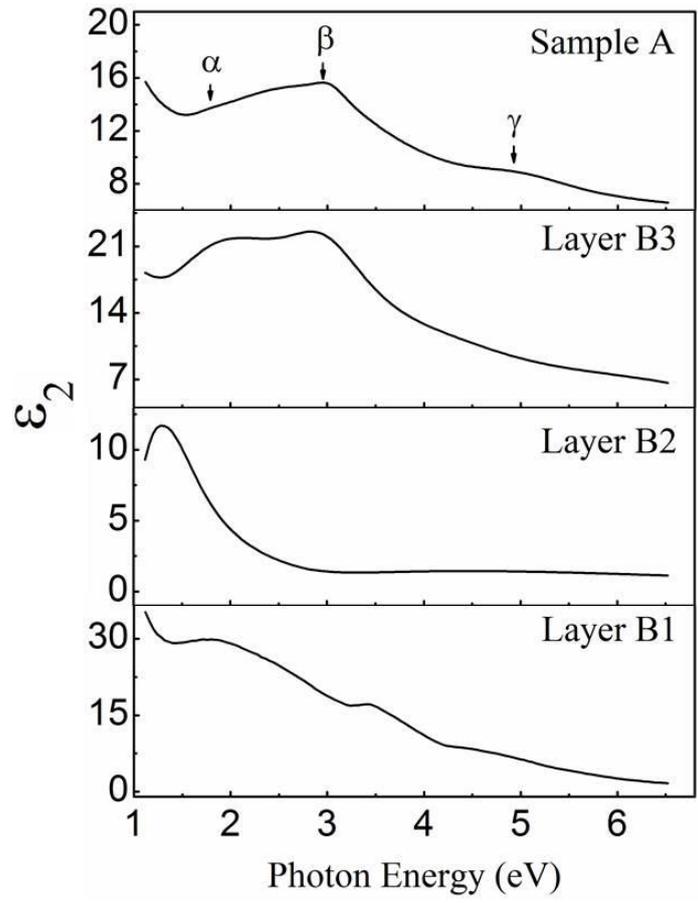

Fig. 10. Imaginary parts of dielectric functions of layers in sample B.